\documentclass[preprint,review,12pt]{elsarticle}

\usepackage{amssymb}

\begin{document}

\begin{frontmatter}



\title{Photometric studies of open star clusters Haffner 11 and Czernik 31.}


\author{D. Bisht\footnote{E-mail: devendrabisht297@gmail.com (D Bisht); rkant@aries.res.in (R K S Yadav) and alokdurgapal@gmail.com (A K Durgapal)}, R. K. S. Yadav$^2$ and A. K. Durgapal$^1$}

\address{$^{1}$Department of Physics, DSB Campus, Kumaun University, Nainital-263002, Uttarakhand, India\\
$^{2}$Aryabhatta Research Institute of Observational Sciences, Manora Peak Nainital 263 002, India; Tel: 0091 05942 235583; Fax No: 0091 05942 233439\\
}
\begin{abstract}

We present the broad band $UBVI$ CCD photometric investigations in 
the region of the two open clusters Haffner 11 and Czernik 31.
The radii of the clusters are determined as 3$^\prime$.5 and 3$^\prime$.0 
for Haffner 11 and Czernik 31 respectively. Using two colour $(U-B)$ 
versus $(B-V)$ diagram we determine the reddening $E(B-V) = 0.50\pm$0.05 
mag and 0.48$\pm$0.05 mag for the cluster Haffner 11 and Czernik 31
respectively. Using 2MASS $JHK_s$ and optical data, we determined 
$E(J-K) = 0.27\pm$0.06 mag and $E(V-K) = 1.37\pm$0.06 for Haffner 11
and $E(J-K) = 0.26\pm$0.08 mag and $E(V-K) = 1.32\pm$0.08 mag for 
Czernik 31. Our analysis indicate normal interstellar extinction 
law in the direction of both the clusters. Distance of the clusters is 
determined as 5.8$\pm$0.5 Kpc for Haffner 11 and 3.2$\pm$0.3 Kpc for 
Czernik 31 by comparing the ZAMS with the CM diagram of the 
clusters. The age of the cluster has been estimated as 800$\pm$100 Myr 
for Haffner 11 and 160$\pm$40 Myr for Czernik 31 using the stellar 
isochrones of metallicity $Z = 0.019$. 

\end{abstract}

\begin{keyword}
Star: evolution -stars: Hertzsprung-Russell and colour-magnitude diagrams
$-$open clusters and associations: individual: Haffner 11 and Czernik 31 


\end{keyword}

\end{frontmatter}


\section{Introduction}
\label{Intro}
Open clusters (OCLs) are located in the Galactic disc. It is important to 
determine their properties and spatial distribution to understand the 
structure and evolution of Milky Way. Therefore, investigation of the OCLs and, 
in particular, the estimation of their physical parameters like age, distance, 
reddening, size and metallicity are very valuable. 

Considering the above we decided to study two open star clusters namely Haffner 11 (Ha 11) 
and Czernik 31 (Cz 31).  Ha 11 was first discovered by Haffner (1957). Further, 
van den Bergh et al. (1975) studied this cluster photographically and derived a diameter 
of 5$^{\prime}$ and richness of type M(moderate). Cz 31 was discovered by 
Czernik (1966) using the charts of Palomar sky atlas. This cluster is classified as a 
Trumpler class IV2p by Ruprecht (1966). 
 
The main purpose of the present investigation is to estimate the fundamental parameters 
of the clusters Ha 11 and Cz 31 using $UBVI$ CCD data. In 
the literature, both the clusters are poorly studied. The basic parameters taken from 
WEBDA and Dias et al. (2002) are listed in Table \ref{para}. Both the clusters are 
located in the third Galactic quadrant and towards the Galactic anticenter direction. 
These objects are very useful to understand the disc subsystem to which the cluster 
belongs. Table \ref{para} shows that the clusters are of intermediate and young age 
objects which can be used to probe the chemical evolution and star formation history 
of the disc.  

In the present study, we provide $UBVI$ CCD photometry for Ha 11 and 
Cz 31 and determine their basic parameters using optical and near-IR 2MASS $JHK_s$ 
data. CCD optical data used in the present analysis is obtained for the first time for 
Cz 31. 

\section{Observations and data reduction}
\label{Obs}
\begin{figure}
\begin{center}
\centering
\vspace{-1cm}
\hbox {
\hspace{-1.0cm}\includegraphics[width=7.5cm]{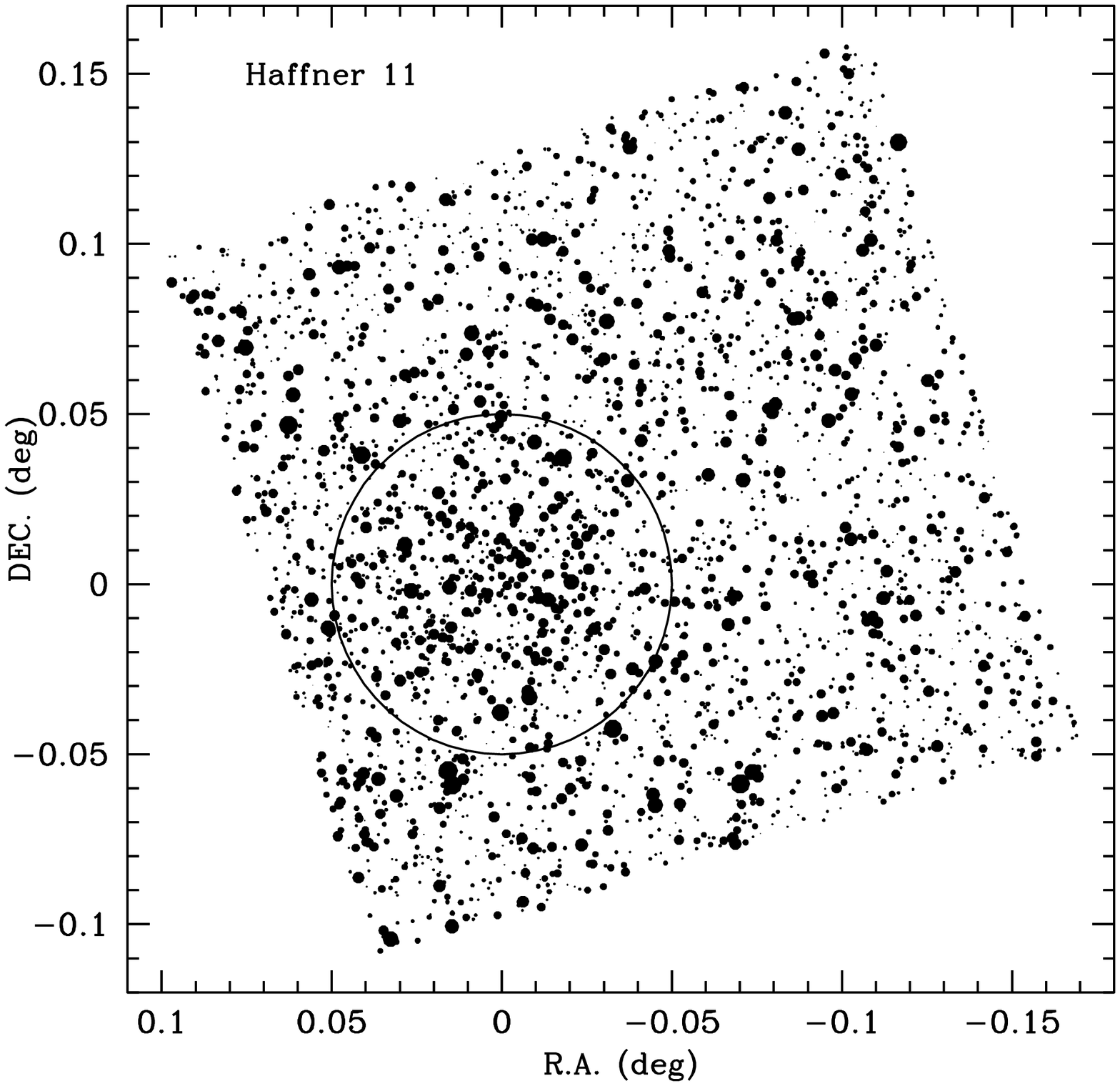}
\includegraphics[width=7.5cm]{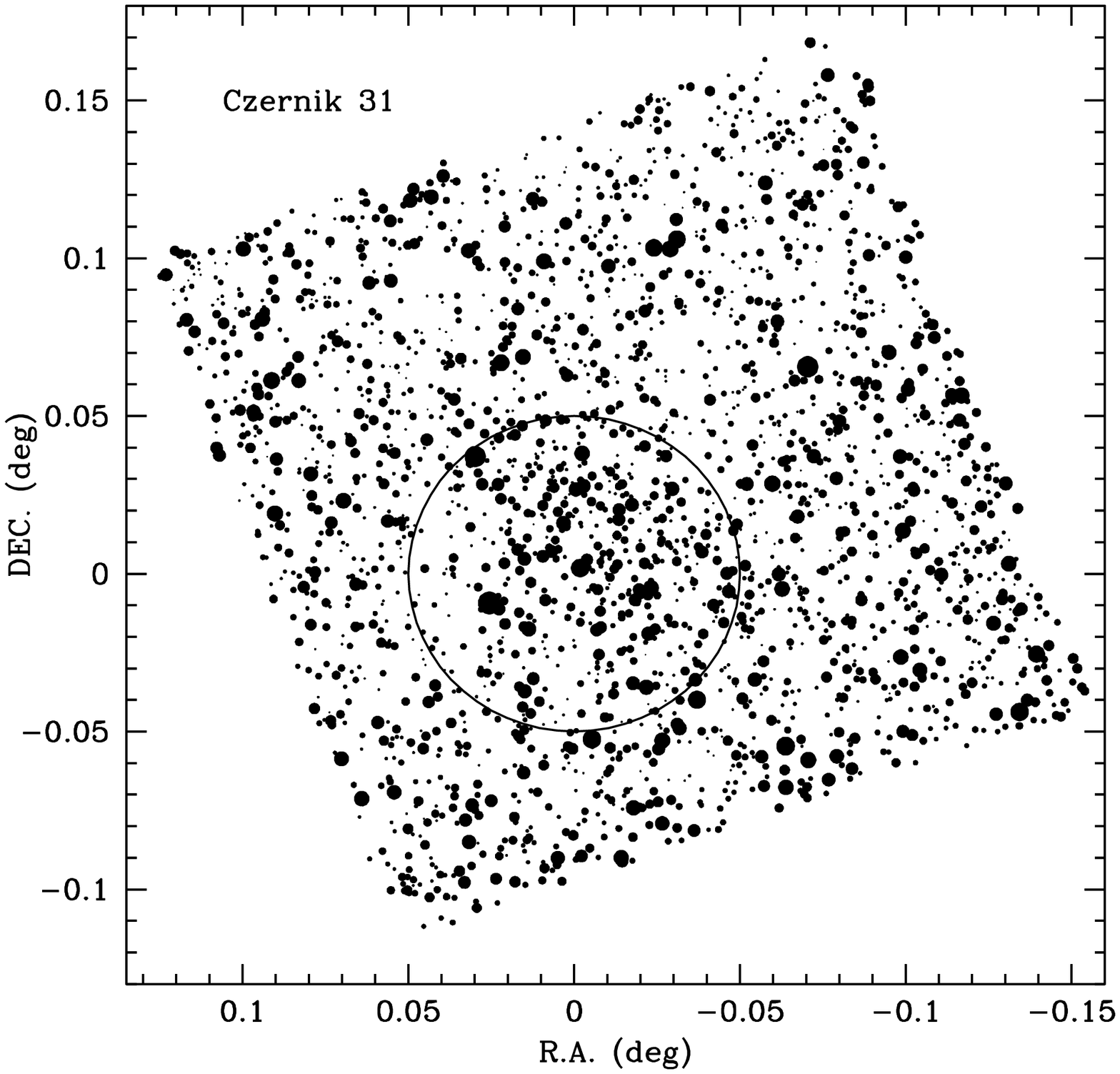}
}
\caption{Finding chart of the stars in the cluster and field regions of
Haffner 11 and Czernik 31. The circle represents the cluster size.}
\label{chart}
\end{center}
\end{figure}

The CCD broad band $UBVI$ images were collected using 
2K$\times$2K CCD system at the $f/13$ Cassegrain focus of the Sumpurnanand 
104-cm telescope located at ARIES, Manora peak, Nainital, India. The used  
CCD has 24 $\mu$m square pixel size, resulting in a scale 
of 0$^{\prime\prime}$.36 pixel$^{-1}$ and a square field of view of 
12$^{\prime}$.6 size. The CCD gain was 10 e$^{-}$/ADU while the read out noise 
was 5.3 e$^{-}$. In order to improve the S/N ratio, the observations were taken 
in 2$\times$2 pixel binning mode. The observations include several exposures 
in each filters. The details of observations are listed in Table \ref{log}. 
Dates of observations together with the filters used and the corresponding 
exposure time are provided in Table \ref{log}. Identification maps of the 
observed regions of the clusters are shown in Fig \ref{chart}.

A number of bias and flat-field frames were taken during the observations. 
Flat-field exposures were made of the twilight sky in each filter. Corrections 
for bias and flat-field were performed using the standard IRAF\footnote{ IRAF is 
distributed by the National Optical Astronomical Observatory which are operated 
by the Association of Universities for Research in Astronomy, under contract with 
the National Science Foundation} procedure. The subsequent data reduction and 
analysis were done using the DAOPHOT software (Stetson (1987), (1992)). The stellar 
photometric routine of DAOPHOT was used for the magnitude determination. These 
magnitudes were calibrated using Landolt (1992) standards. The details of data
reductions are described in the previous papers (Pandey et al. (1997); Durgapal et 
al. (1997, 2001)). The instrumental magnitudes were obtained through quadratically 
varying Point Spread Function (PSF). For each filter, the stars have been aligned to 
that of a reference frame, the deepest one. The average instrumental magnitude was 
derived in each filter. 

We observed the standard field PG 0942$+$051 and SA 98 (Landolt 1992) 
in $U,B,V$ and $I$ for calibrating the observations of Ha 11 and Cz 31. 
The standard stars used in the calibrations have brightness and colours 
range 11.93 $\le V \le$ 15.67 and $-0.29 < (B-V) < 2.19$ respectively, 
thus covering the bulk of the cluster stars. For the extinction coefficients 
we assumed the typical values for the ARIES site.

\begin{figure}
\begin{center}
\includegraphics[width=8.0cm, height=7.0cm]{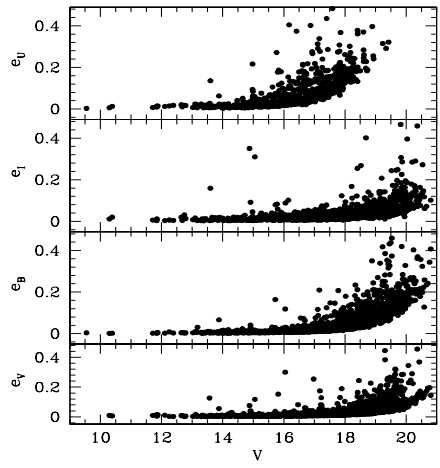}
\caption{Photometric errors in $U, B, V$ and $I$ magnitudes against $V$ magnitude.} 
\label{err_v}
\end{center}
\end{figure}

For translating the instrumental magnitude to the standard 
magnitude, the calibration equations  are as follows:

\begin{center}   

   $u=U+Z_{U}+C_{U}(U-B)+k_{U}X$
   
   $b=B+Z_{B}+C_{B}(B-V)+k_{B}X$
   
   $v=V+Z_{V}+C_{V}(B-V)+k_{V}X$
   
   $i=I+Z_{I}+C_{I}(V-I)+k_{I}X$

\end{center}

where $u, b, v$ and $i$ are the aperture instrumental magnitudes and $U, B, V$ 
and $I$ are the standard magnitudes and $X$ is the airmass. The color 
coefficients ($C$) and zeropoints ($Z$) for different filters are listed in 
Table \ref{std}. 
The errors in zero points and colour coefficients are $\sim$ 0.01 mag for $B, V$ 
and $I$ filters. The internal errors in magnitude derived from DAOPHOT are 
plotted against $V$ magnitude in Fig. \ref{err_v}. This figure shows that photometric 
error is $\le$ 0.01 mag at $V\sim19^{th}$ mag for $B, V$ and $I$ filters while error is 
$\le$ 0.05 mag for $U$ filter at $V\sim17^{th}$ mag. Photometric global 
(DAOPHOT+Calibration) errors 
have also been determined by following Patat et al. (2001) and are listed in 
Table \ref{std_err}. For the $V$ filter, the errors are 0.05 and 0.06 at 
$V\sim$ 17.0 and 20.0 mag, respectively. The final photometric data are 
available in electronic form at the WEBDA site 
\footnote{\it http://obswww.unige.ch/webda/} and also from the authors.

The X and Y coordinates of the stars in the observed region of the clusters 
have been converted to right ascension (RA) and declination (DEC) of J2000. In order to 
obtain an astrometric solution we use the SkyCat tool and Guide Star Catalogue 
v2 (GSC-2) at the European Southern Observatory. This way we considered 100 
bright stars per field for which we have both celestial coordinates on the GSC-2 
and the corresponding pixel coordinates. By using CCMAP and CCTRAN in IRAF, we 
estimated the transformation and compute the individual celestial coordinates 
for all the detected stars. The transformation have an rms value of about 0.10 
arcsec for RA and DEC.  

\section{Analysis of the data}

\subsection{Cluster radius and radial stellar surface density}
\label{rad}
\begin{figure}
\hspace{2cm}\includegraphics[width=8.5cm]{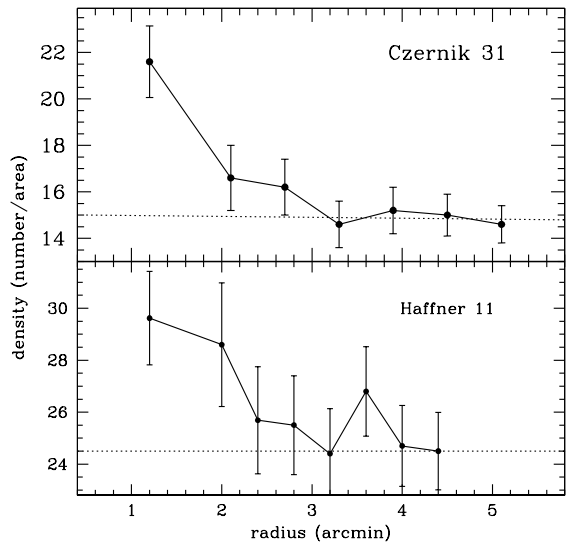}
\caption{Surface density distribution of stars in the field of the cluster 
Haffner 11 and Czernik 31. Errors are determined from sampling 
statistics(=$\frac{1}{\sqrt{N}}$ where $N$ is the number of stars used in 
the density estimation at that point). Dotted line represents the level of field 
stars density.}  
\label{dens}
\end{figure}

To determine the radius of the cluster, we derive the surface stellar density 
by performing star counts in concentric rings around the cluster centre listed in 
Table \ref{para}, and then divided by their respective areas. In Fig. \ref{dens} 
we show the density profile for the cluster Ha 11 and Cz 31. The radial 
density profile of the cluster Ha 11 shows a flattening around 
$r\sim$ 3.5 arcmin and started merging with background stellar density. 
Therefore, we consider 3.5 arcmin as the cluster radius. Our estimate of 
radius is more than the value listed in Table \ref{para}. The radial density 
profile for Cz 31 is decreasing smoothly and $\sim$ 3.0 arcmin, it is mixing 
with field stars density. We consider 3.0 arcmin radius for Cz 31 which is larger 
than the value listed in Table \ref{para}.

\subsection{Colour-magnitude diagrams}
\label{cd}

\begin{figure}
\begin{center}
\hbox{
\hspace{-1.0cm}\includegraphics[width=7.5cm]{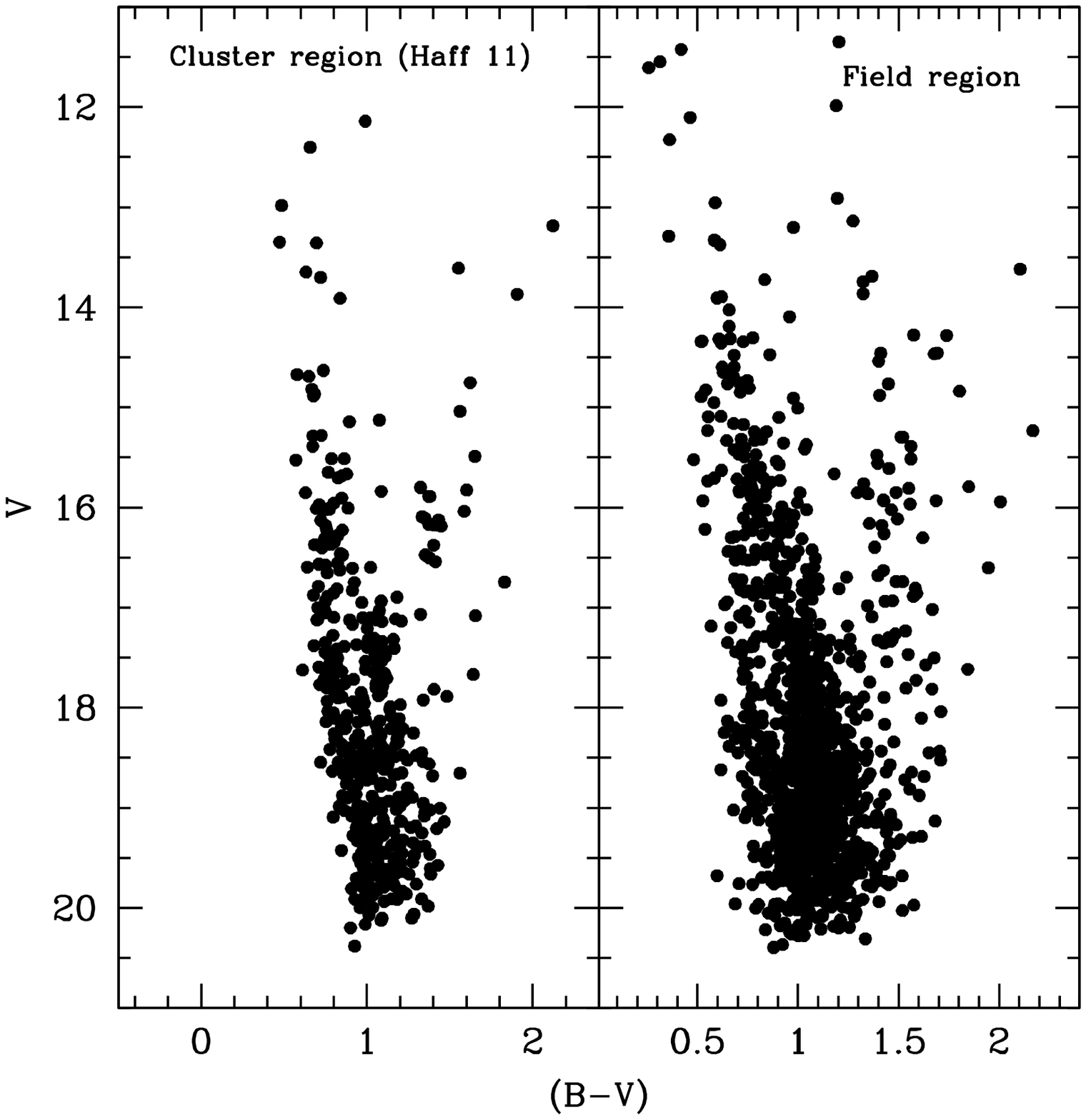}
\includegraphics[width=7.5cm]{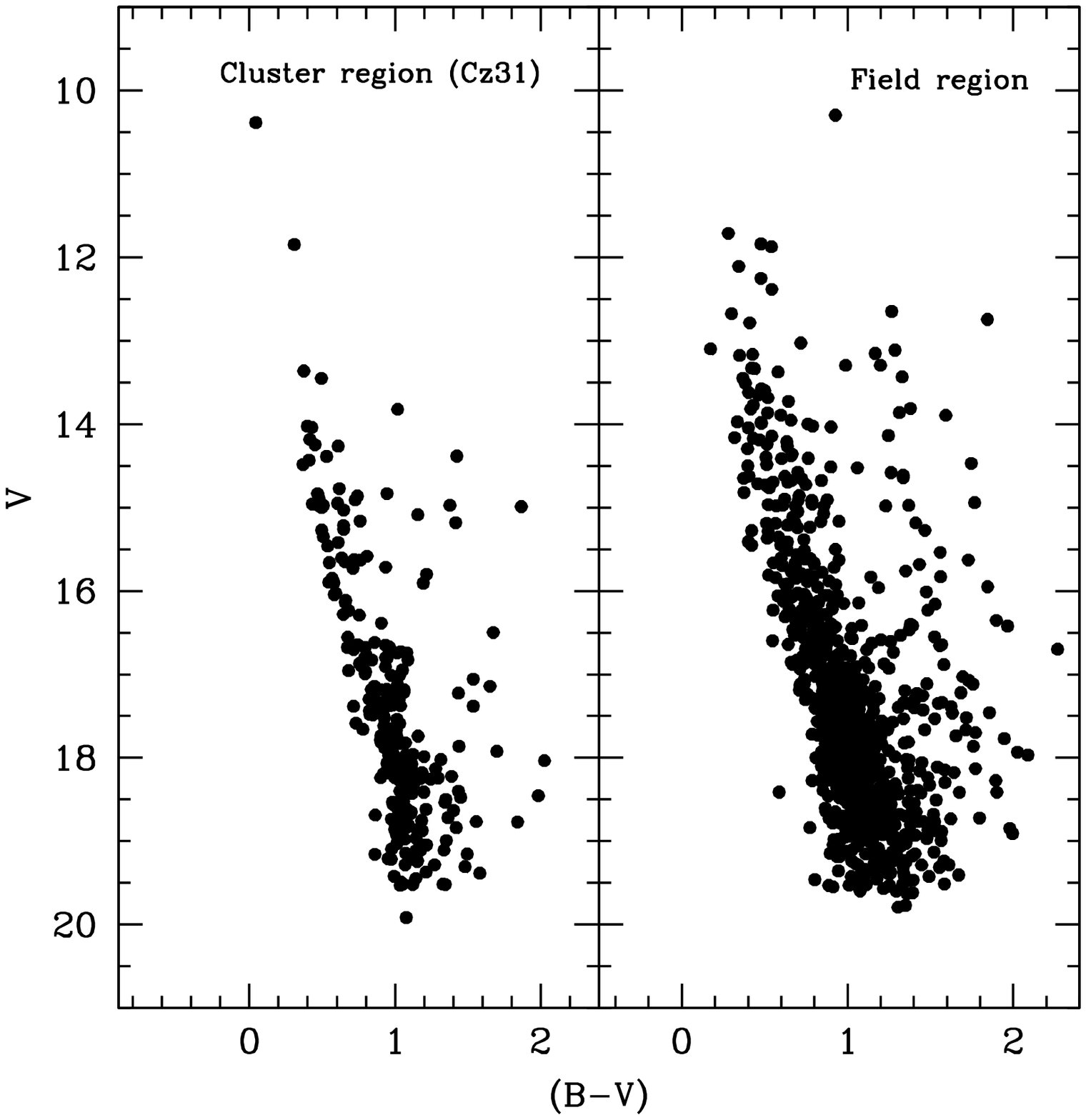}
}
\caption{ The $V, (B-V)$ CMD for the cluster Haffner 11 and Czernik 31 using 
stars within the cluster radius. Stars outside the cluster radius are also 
plotted as field region stars.}
\label{ddd}
\end{center} 
\end{figure}

The $V, (B-V)$ CMDs of the cluster and field region is shown in Fig. \ref{ddd} 
for the cluster Ha 11 and Cz 31. Stars falling within the cluster 
radius are considered as cluster region stars while those outside the 
radius are assumed as field region stars. To get the clear features in the
CMD, we consider the stars as cluster member within cluster radius. Field 
region CMDs shown in Fig. \ref{ddd} are clearly dominated by foreground/background 
stars.\\

{\bf Haffner 11}: The $V, (B-V)$ CMD of this cluster show a main-sequence (MS) extending 
from $V\sim16$ mag, where the turn-off is located, down to $V\sim18$ mag. 
After $V\sim18$ mag, field stars are clearly dominating 
and MS is merging in the field stars. Few red giant stars are also visible 
around $V\sim16$ and $(B-V)\sim1.5$ mag in the CMD. The morphology of the CMD 
indicate that it is a typical intermediate age open cluster. 

{\bf Czernik 31}: A close inspection of the CMD exhibits a poorly 
populated MS extending from $V\sim12$ mag down to $V\sim17$ mag. The 
MS fainter than $V\sim17$ mag has more scatter and field star contamination 
is also more evident. Because of that it 
is hard to separate the cluster members from field stars only on the 
basis of closeness to the main populated area of the CMD. The CMD of this cluster 
looks like a poorly populated young open star cluster.

\subsection{Colour-colour diagram}
\label{red}
\begin{figure}
\begin{center}
\includegraphics[width=9.5cm]{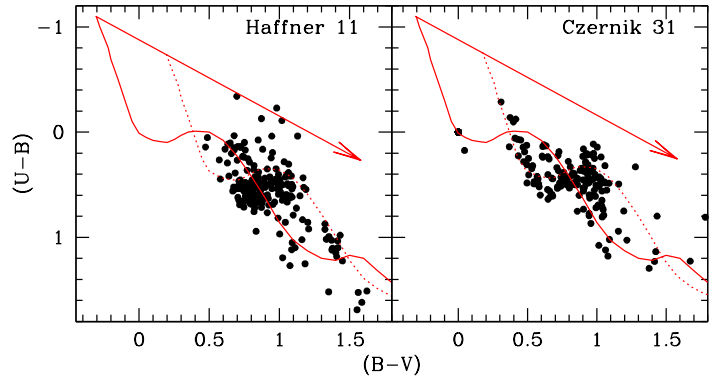}
\vspace{-2.0cm}
\caption{ The $(U-B)$ versus $(B-V)$ colour-colour diagram for the clusters 
under study. The continuous curve represents locus of Schmidt-Kaler's (1982) 
ZAMS for solar metallicity. The dashed lines are the same ZAMS shifted by the 
values given in the text. The solid arrow indicates the reddening vector.}
\label{cc}
\end{center}
\end{figure}

To determine the reddening of the clusters we plot $(U-B)$ versus $(B-V)$ colour-colour 
diagrams in Fig \ref{cc} using probable members of the cluster. The 
intrinsic zero-age main-sequence (ZAMS) given by Schmidt-Kaler (1982) is 
shown by the solid line, whereas the dashed line is the same ZAMS which is 
shifted by assuming the slope of reddening $E(U-B)/E(B-V)$ as 0.72. 
The ZAMS shifted to the MS stars provides a mean value of $E(B-V)=0.50\pm0.05$ 
mag for Ha 11 and $E(B-V)=0.48\pm0.05$ for Cz 31. Our derived values of 
reddening is in agreement with the value listed in Table \ref{para} for Ha 11.
Bica et al. (2005) estimated the reddening $E(B-V)=0.06\pm0.03$ for Cz 31 
using 2MASS $JHK_s$ photometry. Our derived value of reddening is much higher 
than Bica et al. (2005). 

\subsection{Interstellar extinction in near-IR}
\label{red_ir1}
\begin{figure}
\centering
\includegraphics[width=9.5cm]{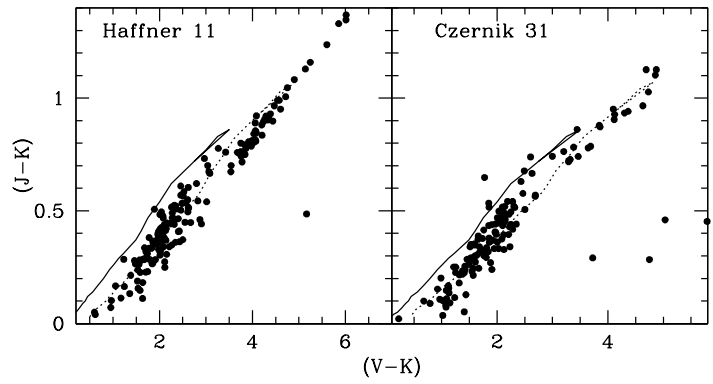}
\vspace{-2.0cm}
\caption{ The plot of $(J-K)$ versus $(V-K)$ colour-colour diagram of the clusters for the
stars within the cluster radius. The solid line is the ZAMS taken from Caldwell et al. (1993). 
The dotted lines are the ZAMS shifted by the values given in the text.} 
\label{red_ir}
\end{figure}

The near-IR data is available for both the clusters in 2MASS catalogue\footnote
{http://vizier.u-strasbg.fr/viz-bin/VizieR?-source=II\%2F246}. The  
$JHK_s$ data in combination with optical data has been used to study the interstellar 
extinction. The $K_s$ magnitudes are converted into $K$ magnitude following 
Persson et al. (1998). The $(J-K)$ versus $(V-K)$ diagram for the clusters under 
study is shown in Fig. \ref{red_ir}. The ZAMS shown by the solid line is taken from 
Caldwell et al. (1993). The same ZAMS is shifted by $E(J-K) = 0.27\pm0.06$ mag 
and $E(V-K) = 1.37\pm0.06$ mag for Ha 11 and $E(J-K) = 0.26\pm0.08$ mag 
and $E(V-K) = 1.32\pm0.08$ mag for Cz 31 and shown by dotted line. The ratios 
$\frac{E(J-K)}{E(V-K)} \sim 0.20\pm0.10$ for both Ha 11 and Cz 31 
is in good agreement with the normal interstellar extinction value of 
0.19 suggested by Cardelli et al. (1989). However, scattering is larger 
due to error in $JHK$ data.

Using the relation $R =$ 1.1$E(V-K)$/$E(B-V)$ given by Whittet et al. (1980) 
we studied about the nature of interstellar extinction law in the direction 
of both the clusters. The values of $R$ derived in this way are 3.0 
for both the cluster Ha 11 and Cz 31. Based on this 
analysis, we can conclude that interstellar extinction law is normal in 
the direction of both the clusters.

\subsection{Age and distance to the clusters}
\label{agedist}
\begin{figure}
\begin{center}
\hbox{
\hspace{-1.4cm}\includegraphics[width=8.0cm]{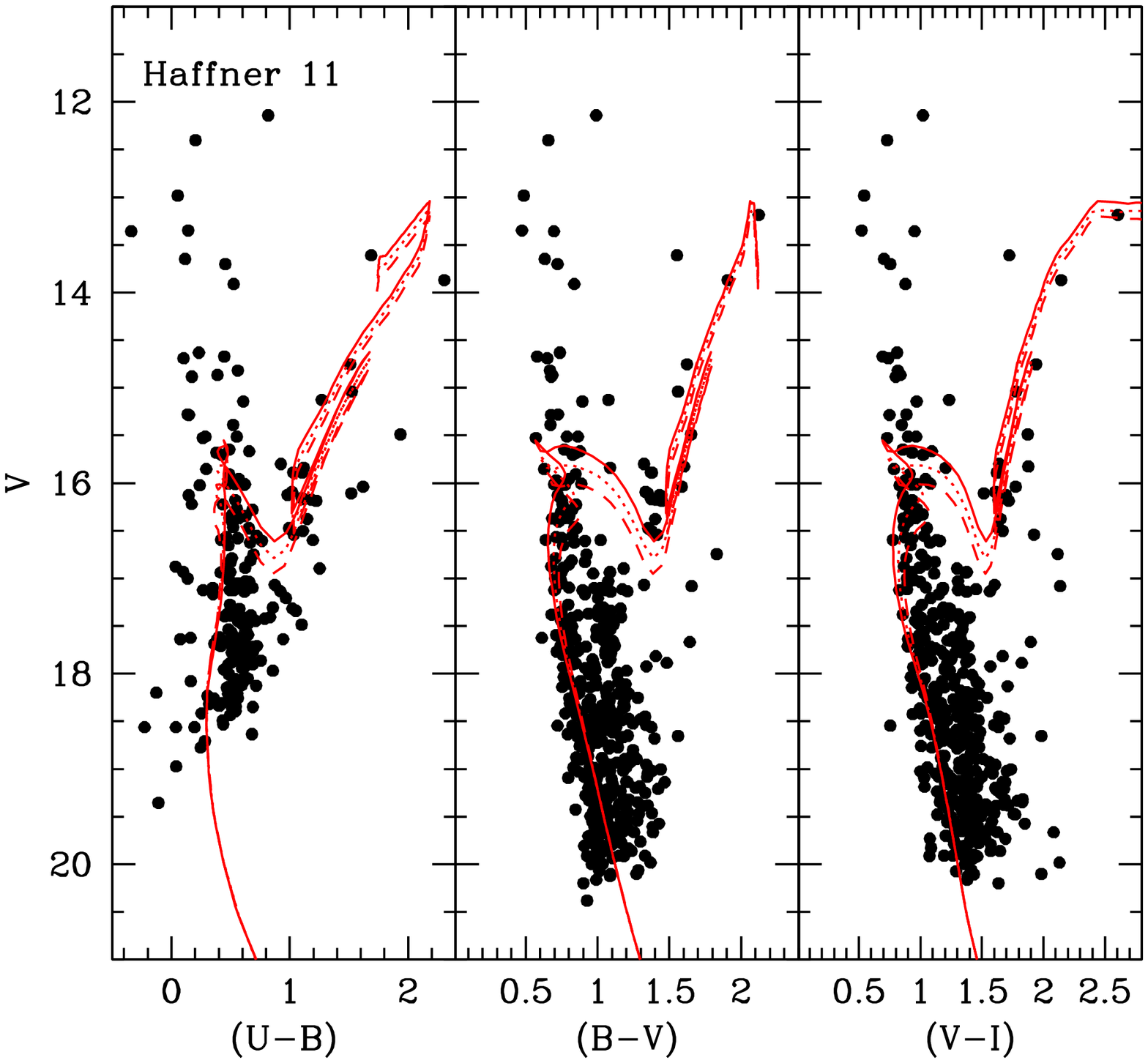}
\includegraphics[width=8.0cm]{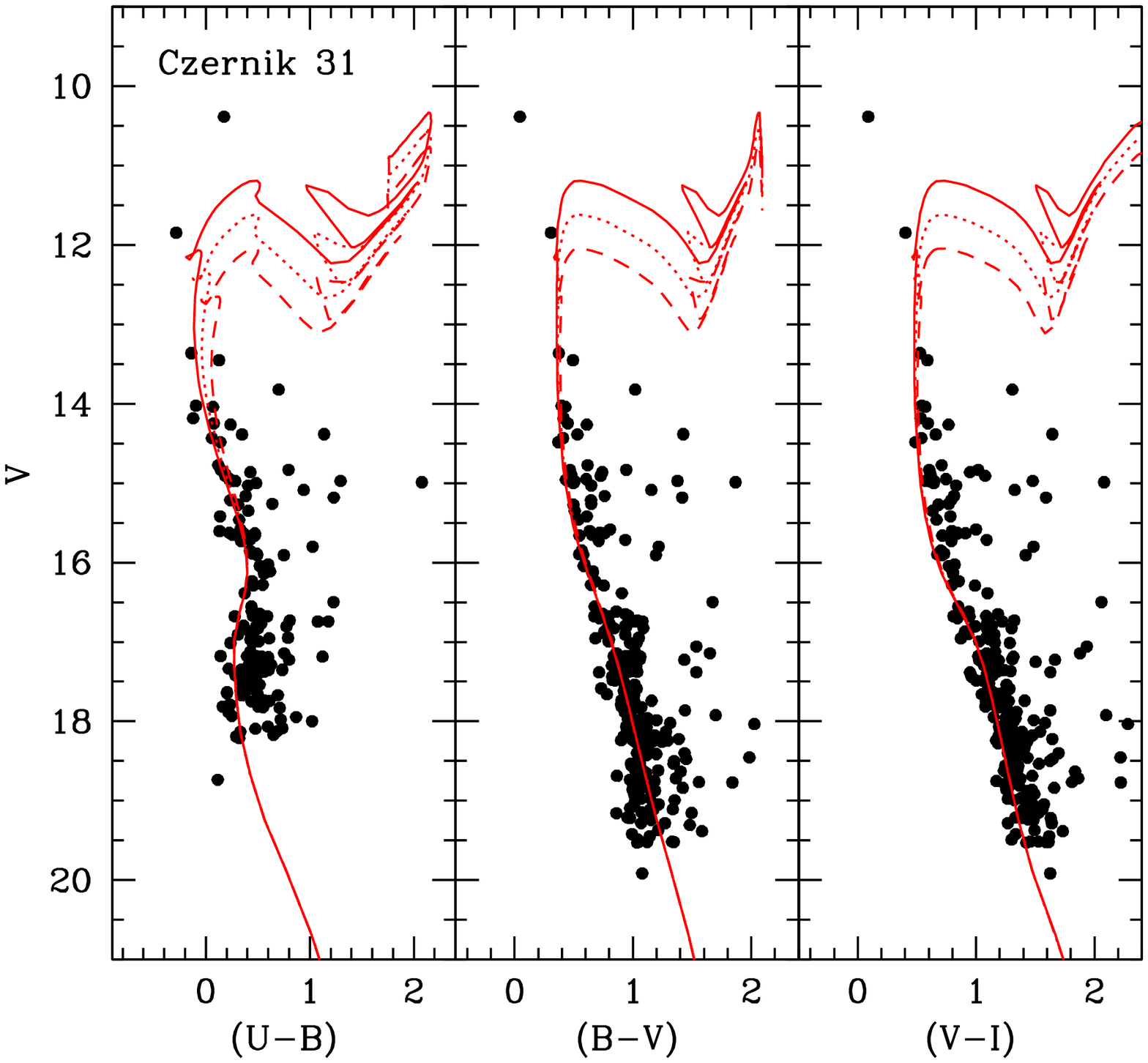}
}
\caption{ The $V, ~(U-B); V, ~(B-V)$ and $V, ~(V-I)$ CMDs for the cluster 
Haffner 11 and Czernik 31 using stars within cluster radius. The different 
lines are the different age isochrones taken from Girardi et al. (2000). Three 
isochrones of different age (log(age)=8.85, 8.90 and 8.95) of metallicity 
$Z=0.019$ are shown in the CMD of Haffner 11. The isochrones of 
log(age)=8.10, 8.20 and 8.30 and $Z=0.019$ are plotted in the CMD of Czernik 31.} 
\label{dist}
\end{center}
\end{figure}

The age of a cluster is determine by comparing the theoretical stellar 
evolutionary isochrones given by Girardi et al. (2000) for $Z=0.019$ with 
the observed CMDs of the clusters shown in Fig. \ref{dist}. The detailed 
shape and position of the different features in the CMD depend mostly on 
age and metallicity and also on reddening and distance. 

We surveyed different age and metallicity isochrones to get the best 
fit isochrones to all the CMD features in $V, ~(U-B); V, ~(B-V)$ and 
$V, ~(V-I)$ diagrams. To get the clear features in the CMDs, we consider 
only those stars which lie within the cluster radius as derived in Section 
\ref{rad}. The uncertainties in age and distance reflect the range 
that allows a reasonable fit to the cluster CMD.

{\bf Haffner 11}: In Fig. \ref{dist}, we show the fitting of isochrones 
to $V, ~(U-B); V, ~(B-V)$ and $V, ~(V-I)$ CMDs using the reddening derived in 
Section \ref{red}. We plotted only those stars which lie within the cluster 
radius derived in Section \ref{rad}. The isochrones of different age 
(log(age)=8.85, 8.90 and 8.95) and $Z=0.019$ have been superimposed on the 
CMDs. The overall fit is good for log(age)=8.90 (middle isochrone). The detailed 
shape of the MS, TO and giant phase are reproduced. 
The best fitting isochrone provides an age of $800\pm100$ Myr. The 
inferred distance modulus $(m-M)=15.36\pm0.2$ mag provide a heliocentric 
distance $5.8\pm0.5$ kpc. The Galactocentric distance is 12.3 kpc, which 
is determined by assuming 8.5 kpc as the distance of the Sun to the 
Galactic center. The distance determined in the present study is in agreement 
with the value listed in Table 1. The Galactocentric coordinates are 
$X=-5.1$ kpc, $Y=2.6$ kpc and $Z=-0.36$ kpc. Using 2MASS $JHK_s$ photometry, 
Bica et al. (2005) derived a heliocentric distance 5.2$\pm$0.2 kpc and age 
890$\pm$150 Myr, which are in agreement with the present estimate of 
distance and age. 

{\bf Czernik 31}: In Fig. \ref{dist}, we superimpose isochrones with $Z=0.019$ 
and for 3 ages (log(age)=8.10, 8.20 and 8.30) in $V, ~(U-B); V, ~(B-V)$ and $V, ~(V-I)$ 
CMDs. The best fit isochrone is found for an age of $160\pm50$ Myr, where 
the associated error has been derived by trying different age isochrones. On 
average, we obtained distance modulus $(m-M)=14.0\pm0.2$ mag. The errors 
have been derived by displacing the best fit isochrone back and forth in 
the distance modulus direction and exploring the acceptable value of distance 
modulus. The estimated distance modulus provides a heliocentric distance 
$3.2\pm0.3$ kpc. The Galactocentric coordinates are $X=-2.7$ kpc, $Y=1.8$ kpc 
and $Z=0.01$ kpc. The Galactocentric distance of the cluster is $10.6$ kpc towards 
the anti-center direction.

\subsection{Near-IR colour-magnitude diagrams}
 
\begin{figure}
\begin{center}
\hbox{
\hspace{-1.8cm}\includegraphics[width=8.5cm]{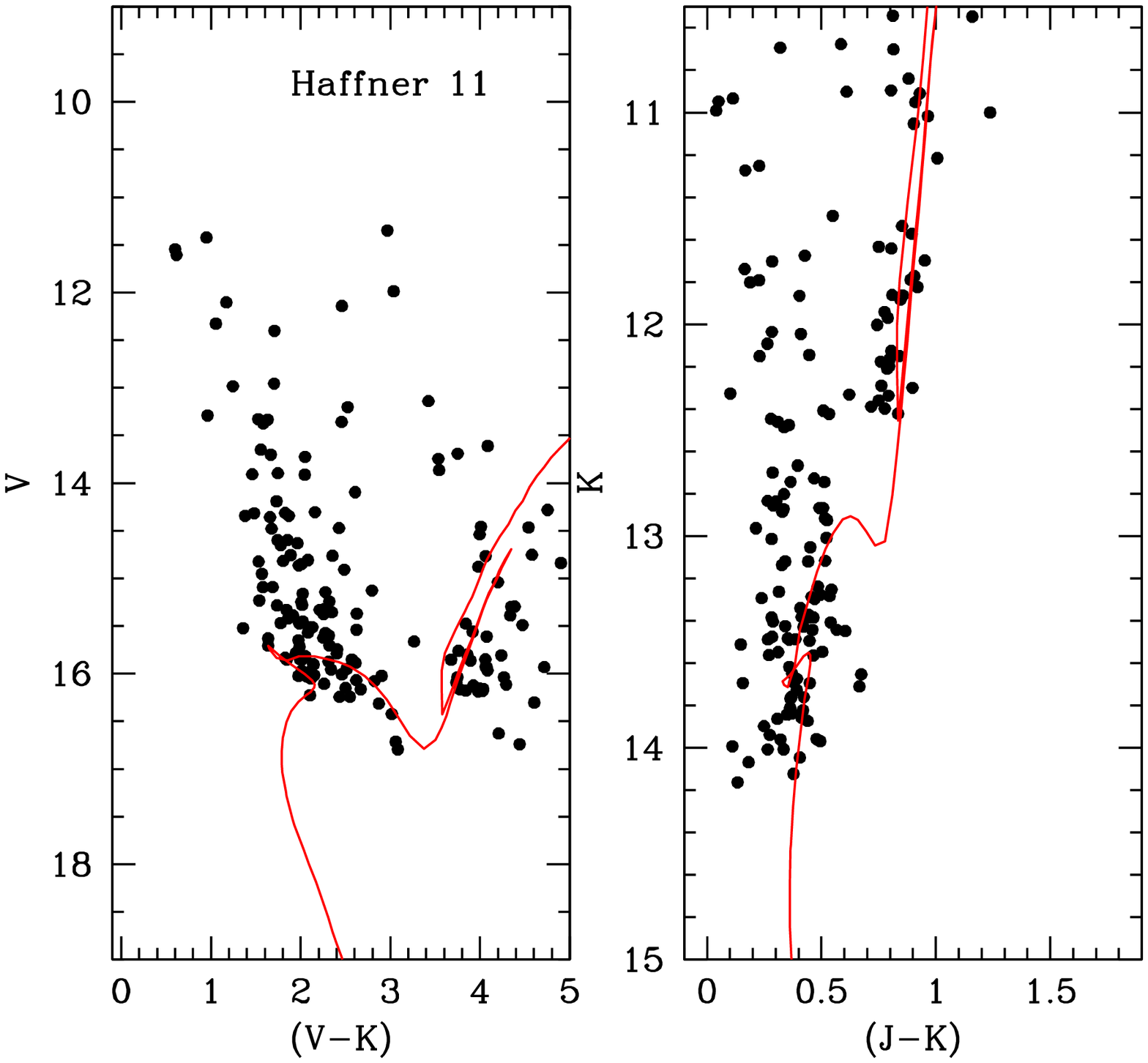}
\includegraphics[width=8.5cm]{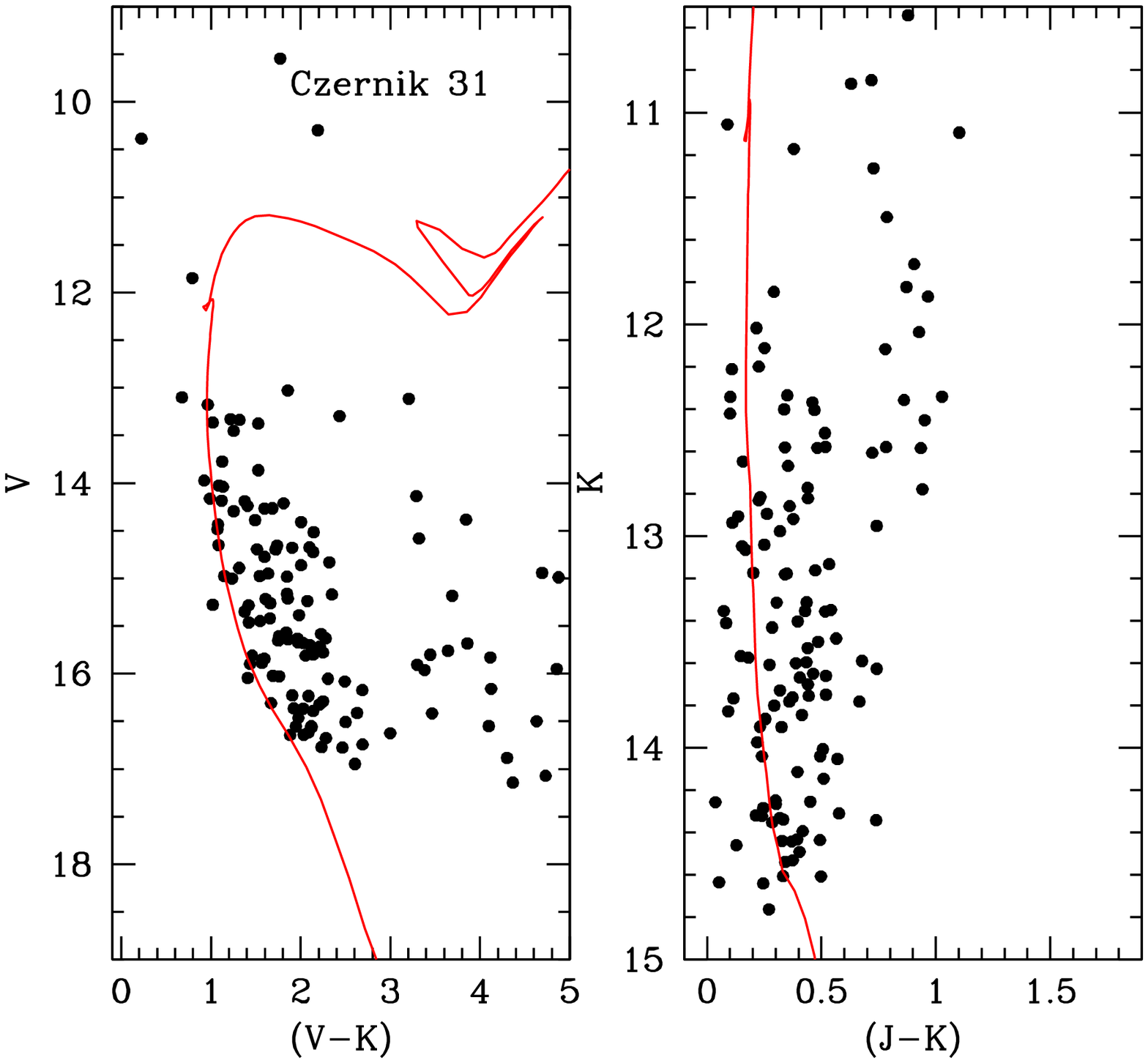}
}
\end{center}
\vspace{-1.0cm}
\caption{ The $V$ versus $(V-K)$ and $K$ versus $(J-K)$ CM diagram of the 
clusters using probable cluster members. The solid curve represent the 
isochrones of log(age) $=$ 8.90 for Haffner 11 and log(age) $=$ 8.20
for Czernik 31 taken from Girardi et al. (2000) for $Z =$ 0.019.}
\label{cmdir}
\end{figure}

Using optical and near-IR data we redetermined distance and age of the 
clusters under study. We plot $V, (V-K)$ and $K, (J-K)$ CMDs for Ha 11 
and Cz 31 in Fig. \ref{cmdir}. The theoretical isochrone given by 
Girardi et al. (2000) for $Z=0.019$ have been overplotted for log(age)=8.9
and 8.1 in the CMDs of Ha 11 and Cz 31 respectively. The apparent distance moduli 
$(m-M)_{V,(V-K)}$ and $(m-M)_{J,(J-K)}$ turn out to be $15.4\pm0.3$ and 
$13.7\pm0.3$ for Ha 11 and $14.0\pm0.3$ and $12.6\pm0.3$ for Cz 31. 
Using the reddening derived in Section \ref{red_ir1}, we derive a distance of 
$5.4\pm0.5$ and $3.3\pm0.3$ kpc for Ha 11 and Cz 31 respectively. 
Both, age and distance determination for the present clusters is in 
agreement with the estimates presented in Section \ref{agedist}.

\section{Conclusions}
\label{con}
We studied two open star clusters Haffner 11 and Czernik 31 using $UBVI$ CCD 
optical and 2MASS $JHK_s$ data. The results are summarized in Table \ref{sum}. The 
main findings of our analysis are given below.

\begin{enumerate}
\item The radii of the clusters are obtained as 3$^\prime$.5 and 3$^\prime$.0
 which corresponds to 5.9 pc and 2.8 pc, respectively, at a distance of the 
cluster for  Ha 11 and Cz 31.

\item From the two colour $(U-B)$ versus $(B-V)$ diagram, we estimated  
$E(B-V)=0.50\pm0.05$ mag for Ha 11 and $0.48\pm0.05$ mag for Cz 31.

\item Distance to the cluster Ha 11 and Cz 31 are determined as
5.8$\pm$0.5 and 3.2$\pm$0.3 Kpc respectively. Age of 800$\pm$100 Myr and 
160$\pm$40 Myr are determined for the clusters Ha 11 and Cz 31 respectively by comparing 
the isochrones of $Z= 0.019$ given by Girardi et al. (2000). These distances 
and ages are supported by the values derived by combining optical and near-IR data. 

\end{enumerate}

\section{ACKNOWLEDGEMENTS}
\label{ACK}
D. Bisht and A. K. Durgapal would like to acknowledge the great 
support from ARIES during the observations and Data reduction. This publication 
made use of data from the 2MASS, which is a joint project of the university of 
Massachusetts and the Infrared Processing and Analysis Center/California Institute 
of Technology, funded by the National Science Foundation. We are also much obliged 
for the use of the NASA Astrophysics Data System, of the SIMBAD data base and of the 
WEBDA open cluster data base.     

~                     
\bibliographystyle{model1a-num-names}
\bibliography{<your-bib-database>}


\clearpage

\clearpage

\clearpage

\clearpage

\clearpage

\clearpage

\clearpage


\clearpage
\begin{table*}
\centering
\caption{Fundamental parameters of the clusters taken from WEBDA and Dias et al.
(2002).}
\vspace{0.5cm} 
\begin{tabular}{ccccccccc}
\hline\hline
Name       & $\alpha_{2000}$  & $\delta_{2000}$ & $l$& $b$& Dia & $E(B-V)$     &
$D$   & log(age)   \\
           & h:m:s            & d:m:s         & (deg) &(deg)  & ($\prime$) &   (mag)        &
 (pc)       &            \\
\hline
Haffner 11 & 07:36:25       & $-$27:43:00       & 242.4 & $-$3.5 &   5     &0.57& 6100 &8.70\\
Czernik 31 & 07:36:59       & $-$20:32:00       & 236.3 & 0.3 &   5     &0.06& 2200&8.25\\
\hline
\end{tabular}
\label{para}
\end{table*}

\clearpage
\begin{table}
\begin{center}
\caption{Log of observations, with dates and exposure times for each passband.}
\vspace{0.5cm}
\begin{tabular}{ccc}
\hline\hline
Band  &Exposure Time &Date\\
&(in seconds)   & \\
\hline\hline
&Haffner 11&\\
$U$&1500$\times$2, 300$\times$2&6$^{th}$ Feb 2010\\
$B$&1200$\times$2, 240$\times$1 &,,\\
$V$&900$\times$2, 180$\times$2 &,,\\
$I$&300$\times$2, 60$\times$1 &,,\\
\hline
&Czernik 31&\\
$U$&1200$\times$1, 240$\times$2&5$^{th}$ March 2010 \\
$B$&900$\times$1, 180$\times$2&,,\\
$V$&600$\times$1, 120$\times$2&,,\\
$I$&300$\times$2, 120$\times$1&,,\\
\hline
\end{tabular}
\label{log}
\end{center}
\end{table}

\clearpage
\begin{table}
\begin{center}
\caption{Derived Standardization coefficients and its errors.}
\vspace{0.5cm}
\begin{tabular}{ccc}
\hline\hline
Filter  &   Colour Coeff. $(C)$ & Zeropoint $(Z)$\\
\hline\hline
&Haffner 11&\\

$U$&$-0.040\pm$0.09&$7.35\pm$0.05\\
$B$&$+0.001\pm$0.01&$5.29\pm$0.01\\
$V$&$-0.104\pm$0.01&$4.99\pm$0.01\\
$I$&$-0.125\pm$0.01&$5.43\pm$0.01\\

&Czernik 31&\\

$U$&$-0.03\pm$0.02&$7.85\pm$0.01\\
$B$&$-0.03\pm$0.01&$5.64\pm$0.01\\
$V$&$-0.06\pm$0.01&$5.21\pm$0.01\\
$I$&$-0.06\pm$0.01&$5.48\pm$0.01\\
\hline
\end{tabular}
\label{std}
\end{center}
\end{table}

\clearpage
\begin{table}
\begin{center}
\caption{The rms global photometric errors as a function of magnitude.}
\vspace{0.5cm}
\begin{tabular}{ccccc}
\hline
$V$&$\sigma_{V}$&$\sigma_{B}$&$\sigma_{I}$&$\sigma_{U}$ \\
\hline
$ 9-11$&$0.03$&$0.05$&$0.03$&$0.05$ \\
$11-12$&$0.05$&$0.05$&$0.03$&$0.05$ \\
$12-13$&$0.05$&$0.05$&$0.05$&$0.06$ \\
$13-14$&$0.05$&$0.05$&$0.05$&$0.06$ \\
$14-15$&$0.05$&$0.05$&$0.05$&$0.06$ \\
$15-16$&$0.05$&$0.05$&$0.05$&$0.08$ \\
$16-17$&$0.05$&$0.05$&$0.05$&$0.11$ \\
$17-18$&$0.06$&$0.06$&$0.06$&$0.13$ \\
\hline
\end{tabular}
\label{std_err}
\end{center}
\end{table}

\clearpage
\begin{table}
\caption{Derived fundamental parameters of the clusters under study. $R_{GC}$ is the 
Galactocentric distance while X, Y and Z are the Galactocentric coordinates 
of the clusters. The coordinate system is such that the Y-axis connects the 
Sun to the Galactic Centre, while the X-axis is perpendicular to that. Y is 
positive towards the Galactic anticentre, and X is positive in the first and 
second Galactic quadrants (Lynga 1982).}

\begin{center}
\small
\begin{tabular}{ccccccccc}
\hline
Name& Radius          & $E(B-V)$ &Distance      & X       & Y      & Z      & $R
_{GC}$ & Age \\
    &(arcmin)         & (mag)    &(kpc)         &(kpc)    &(kpc)   &(kpc)     & (kpc) &(Myr) \\  
\hline
Haffner 11 & 3.5 & $0.50\pm0.05$ & $5.8\pm0.5$ & -5.1 &  2.6 & -0.36 & 
12.3 & $800\pm100$ \\
Czernik 31 & 3.0 & $0.48\pm0.05$ & $3.2\pm0.3$ & -2.7 & 1.8 & 0.01 & 10.6
& $160\pm40$ \\
\hline\hline
\end{tabular}
\label{sum}
\end{center}
\end{table}

\end{document}